\documentclass[a4paper, 11pt]{article}
\pdfoutput=1

\usepackage{amsfonts, amsmath, amssymb, amsthm, amscd, setspace, subfigure}
\usepackage{graphicx}
\usepackage{jheppub}
\usepackage[all,cmtip]{xy}
\newcommand{\norm}[1]{\ensuremath{\left |\left | #1 \right | \right | }}

\newcommand{\ok}{}
\newcommand{\x}{\alpha }
\newcommand{\xx}{\hat\alpha }
\renewcommand{\S}{S}
\renewcommand{\SS}{\Sigma}
\renewcommand{\O}{W}
\newcommand{\OO}{\Omega}

\newcommand{\oo}{\omega}

\makeatletter
\def\@fpheader{\relax}
\makeatother

\title{Harmonic forms on ALF gravitational instantons}
\author{Guido Franchetti}
\affiliation{Maxwell Institute for Mathematical Sciences and Department of Mathematics,\\
Heriot-Watt University, Edinburgh EH14 4AS, UK.}

\emailAdd{G.Franchetti@hw.ac.uk}

\abstract{We study the space of square-integrable harmonic forms over ALF gravitational instantons of type $A _{ K -1 } $ and of type $D _K $. We first calculate its dimension making use of a result by Hausel, Hunsicker and Mazzeo which relates the Hodge cohomology of a gravitational instanton $M$ to the singular cohomology of a particular compactification $X _M $ of $M$. We then exhibit an explicit basis, exact for $A _{ K -1 } $ and approximate for $D _K $, and interpret geometrically the relations between $M$, $X _M $ and their cohomologies.}


\arxivnumber{}

\begin{document}
\maketitle
\flushbottom

\section{Introduction}
The aim of this paper is to investigate the Hodge cohomology, that is the cohomology of square-integrable harmonic forms, of ALF gravitational instantons.

ALF gravitational instantons arise in Euclidean approaches to quantum gravity \cite{Hawking:1976jb}, as moduli space of monopoles \cite{Cherkis:1998cz}, as quantum moduli spaces of supersymmetric gauge theories \cite{Seiberg:307287}, and as compactifications in string theory \cite{Sen:1997tw}. In the context of the geometric models of matter framework \cite{Atiyah:2012hw}, which aims to model static particles via Riemannian 4-manifolds, their r\^ole as models for multi-particle systems has been considered in \cite{Franchetti:2013ib}. They come in two infinite families: of type $A _{ K -1 } $ and of type $D _K $.

Square-integrable ($L^2$) harmonic forms are of natural interest e.g.~in relation with various duality conjectures arising in string theory and as electric fields associated to charged particles in the geometric models of matter framework.

On a compact  orientable Riemannian manifold  de Rham cohomology is isomorphic to Hodge cohomology, but on a non-compact manifold there is generally no such correspondence.
However if a  Riemannian manifold $M$  has a particular asymptotic behaviour (metric of fibred boundary type or of cusp type), which includes that of ALF gravitational instantons,  then there is a relation between the Hodge cohomology of $M$ and the ordinary cohomology of a particular compactification $X _M $ of $M$ \cite{Hausel:2004vv}. We will refer to $X _M $ as the Hausel-Hunsicker-Mazzeo (HHM) compactification of $M$. This work originated as an attempt to elucidate the correspondence between $M$ and $X _M $ in the particular case of ALF gravitational instantons.

The plan of the paper is as follows: in section \ref{dimhf} we recall the topological properties of an ALF gravitational instanton $M$ and of its HHM compactification $X _M $ which are needed in order to calculate the dimension of $L ^2 \mathcal{H} ^p (M) $, the space of square-integrable harmonic $p$-forms on $M$.
In section \ref{hfbasis} we describe the metric properties of ALF gravitational instantons and exhibit an explicit basis of $L ^2 \mathcal{H} ^2 (M) $, the only non-trivial Hodge cohomology group. The results that we obtain are exact in the case of ALF 
$A _{ K -1 } $, approximate in the case of ALF $D _K $ where we rely on an asymptotic approximation of the true metric.

\section{The dimension of $L ^2 \mathcal{H} ^p (M) $}
\label{dimhf} 
A (non-compact) gravitational instanton is a complete hyperk\"ahler 4-manifold with curvature tensor decaying at infinity.
An ALF (short for asymptotically locally flat) gravitational instanton is, outside a compact set, 
the total space of a circle fibration over $\mathbb{R}  ^3 $ or $\mathbb{R}  ^3 / \mathbb{Z}   _2 $ with  fibres of asymptotically constant length. Two infinite families
of ALF gravitational instantons are known: of type $A _{ K -1 } $,  $\mathbb{Z}  \ni K \geq 1 $ and of type $D _K $, $\mathbb{Z}  \ni K \geq 0 $. ALF $A _{ K -1 } $ is also known as multi Taub-NUT with $K$ NUTs  --- the number of NUTs being  the  reason why we prefer to work with $A _{ K -1 }$ rather than with $A _K $.

ALF gravitational instantons of type $A _{K -1 }$, $K \geq 2 $, are topologically the minimal resolution of the Kleinian singularity $\mathbb{C}  ^2 / \mathbb{Z}  _K $. ALF gravitational instantons of type $D _K $, $K \geq 3 $, are the minimal resolution of the Kleinian singularity $\mathbb{C}  ^2 /D ^\ast _{ K -2 } $, where $D ^\ast _{ K} $ denotes the binary dihedral group of order $4 K $.
As a consequence, ALF $A _{ K -1 } $, $K \geq 2 $  (ALF $D _K $, $K \geq 3 $), retracts onto a configuration of 2-spheres intersecting according to the Lie algebra $A _{ K -1 } $ ($D_K$).

Small values of $K$ need a separate description: ALF $A _0 $ is topologically  $\mathbb{C}  ^2 $.
ALF $D _0 $, the moduli space of centered $SU (2) $ monopoles of charge 2 (or non-simply connected Atiyah-Hitchin manifold), retracts onto the real projective plane $P _2 (\mathbb{R}  )$. ALF $ D _1 $, the 1-parameter family of deformations of the double cover of $D _0 $ (or simply connected Atiyah-Hitchin manifold) discovered by Dancer \cite{Dancer:1993vu}, retracts onto a 2-sphere. ALF $D _2 $, the minimal resolution of singularities of $(\mathbb{R}  ^3 \times S ^1 )/ \mathbb{Z}  _2 $,  $\mathbb{Z}  _2 $ acting with two fixed points, retracts onto a configuration of two 2-spheres intersecting according to the Lie algebra $D _2 \simeq A _1 \times A _1 $.

As a consequence of the topological properties mentioned above, the de Rham cohomology of ALF gravitational instantons is given by
\begin{equation}
\label{drc} 
H ^p _{ \mathrm{dR}~}  (M _{ K } ) =\begin{cases}
\mathbb{R}  &\text{if $p=0 $},\\
\mathbb{R}  ^{ K } &\text{if  $p=2 $},\\
0 &\text{otherwise}, \end{cases} 
\end{equation} 
where $ M _K $ stands for either $A _K $ or $D _K $ and  $\mathbb{R}  ^0  $ denotes the trivial vector space.

For both families 
$M \setminus C $, with $C$ a suitable compact set, is a circle fibration over $N = I \times \Sigma $, with $I$  an open interval and $\SS$ a smooth 2-manifold. Let us parametrise the interval $I$ with a coordinate $r$ and denote by $\SS _r =\{r\} \times \SS $. Then the HHM compactification $X _M  $ of an ALF gravitational instanton $M$ is obtained by collapsing the fibre above each point of $\SS _r $ in the limit $r \rightarrow \infty $. Therefore $M = X _M \setminus \SS _\infty $ with $\SS _\infty $, which is diffeomorphic to $\SS $, playing the r\^ole of the spatial infinity of $M$. It is known \cite{Hausel:2004vv,Etesi:2006bc} that $X _M $ is a closed smooth manifold.\footnote{The HHM compactification can be defined in a wider context, but it is generally only a stratified space.} 

By a result (corollary 1) in \cite{Hausel:2004vv}   we have
\begin{equation}
L ^2 \mathcal{H} ^p (M) = \begin{cases}
H ^p (X _M , \SS _\infty; \mathbb{R}   ) & \text{if $p \leq 1$},\\
H ^2 (X_M; \mathbb{R}  ) & \text{if $p =2 $},\\
H ^p (M; \mathbb{R}  ) &\text{if $p \geq 3 $},
\end{cases} 
\end{equation} 
where $H ^p  (A; \mathbb{R}  )$, $ H ^p (A,B; \mathbb{R}  ) $ denote the singular and relative cohomology of $A$ with real coefficients.
For a smooth manifold de Rham cohomology and singular cohomology over $\mathbb{R} $ are isomorphic so we can work with the former. Since for any ALF gravitational instanton $H ^p _{ \mathrm{dR} } (M) = 0 $ if $p \geq 3 $  and $H ^p (X _M , \SS _\infty; \mathbb{R}   )=0 $ for $p =0,1 $, 
\begin{equation}
L ^2 \mathcal{H} ^p (M) = \begin{cases}
H ^2 _{ \mathrm{dR}} (X_M) & \text{if $p =2 $},\\
0 &\text{otherwise}.
\end{cases} 
\end{equation} 

In order to compute the dimension of $L ^2 \mathcal{H} ^2 (M ) $ it is convenient to use a Mayer-Vietoris sequence over the open sets $U = X _M \setminus \SS _\infty  $, $V $ an open neighbourhood of $\SS _\infty $ in $X _M $. Since for $r>0 $ $\SS _r $ is  the base of a circle fibration while for $ r = \infty  $ the fibres have collapsed to zero size, $V$ is a disk bundle over $\SS _\infty $, homotopically equivalent to $\SS _\infty $. The intersection $ U \cap V $ retracts onto a hypersurface of large $r$.

In the case of $A _{ K -1 } $,  $\SS _r $ is diffeomorphic to the  2-sphere $S ^2 $, and a hypersurface of large $r$ has the topology of $S ^3 $ if $K =1 $ and of the lens space $L(K,1) $ if $K >1 $, see appendix \ref{ga}. A Mayer-Vietoris sequence then gives
\begin{equation}
\label{homak} 
H ^p  _{\mathrm{dR}}  (X _{ A _{K-1}}  )=
\begin{cases} 
\mathbb{R}    &\text{if $p =0,4$},\\
\mathbb{R}  ^{K} & \text{if $p =2 $},\\ 
0 &\text{otherwise},
\end{cases} 
\end{equation} 
hence 
\begin{equation}
\mathrm{dim} \left( L ^2 \mathcal{H} ^2 (A _{ K -1 } ) \right) = K .
\end{equation}

In the case of $D _{ K } $,  $\SS _r $ is diffeomorphic to the real projective plane  $P _2 (\mathbb{R}  )$. A hypersurface of large $r$ has the topology of $S^3/D^*_{K-2} $ if $K \geq 3 $, of $S^3/D^*_{2} $ if $K =0 $, of $S^3/D^*_{1} $ if $K =1 $  and of  $(S^ 2 \times S ^1)/ \mathbb{Z}  _2  $ if $K =2 $, see appendix \ref{ga}. A Mayer-Vietoris sequence then gives
\begin{equation}
\label{homdk} 
H ^p _{ \mathrm{dR} } (X _{ D _K } )
=\begin{cases} 
\mathbb{R}  &\text{if $p =0,4$},\\
\mathbb{R}  ^{K} & \text{if $p =2 $},\\ 
0 &\text{otherwise},
\end{cases} 
\end{equation} 
hence 
\begin{equation}
\label{dimdk} 
\mathrm{dim} \left( L ^2 \mathcal{H} ^2 (D _{ K } ) \right) = K .
\end{equation} 
Note how, differently from the case of $A _{ K -1 } $, $\SS _\infty $ does not contribute to the middle dimension cohomology of $X _{ D _K } $.

While $H ^2 _{ \mathrm{dR}} (X_M) $ was all we needed in order to calculate the dimension of $L ^2 \mathcal{H} ^2 (M) $, we should point out that the topology of $X _{ A _{ K -1 } } $ is known:  $X_{A _{ K -1} } $  is homeomorphic to the connected sum of a number of copies of $P _2 (\mathbb{C}  ) $ (with our choice of orientation) equal to $\mathrm{dim} \left( L ^2 \mathcal{H} ^2 (A _{ K -1 } ) \right)  =K $ \cite{Etesi:2006bc}. 
In order to prove this result it is enough to show that the intersection matrix of $X _{ A _{ K -1 } } $ is definite and diagonal. Since  $X _{ A _{ K -1 } } $ is smooth closed oriented and simply connected, the result follows from a theorem by Freedman \cite{Anonymous:2NT017mt}.\footnote{Simply connectedness of $X _{ A _{ K -1 }} $, $X _{ D _K } $ can be shown by applying van Kampen's theorem to the open sets $U$, $V$ used in the Mayer-Vietoris sequence.}
In section \ref{hfbasis} we will exhibit a basis $\{ [\SS ^I] \}$ of $H _2 (X _{ A _{ K -1 }}; \mathbb{Z}  ) $ from which the intersection matrix of $X _{ A _{ K -1 }} $ can be readily calculated and shown to have the fore mentioned properties.

\section{A basis of $L ^2 \mathcal{H} ^2 (M) $}
\label{hfbasis} 
While the computation of $\mathrm{dim}( L ^2 \mathcal{H} ^2  (M) )$ relied only on the topology of an ALF gravitational instanton and of its HHM compactification, in order to exhibit a basis we need to take the metric into account.
ALF gravitational instantons are geodesically complete hence any square-integrable harmonic form is both closed and co-closed. 

Let us start with the $A _{ K -1 }$ family. By a result of Hitchin \cite{Hitchin:1999db}, any square-integrable harmonic 2-form on $A _{ K -1 } $ is anti self-dual with respect to the orientation induced by the hyperk\"ahler structure. In order to follow the conventions used in \cite{Atiyah:2012hw,Franchetti:2013ib} we will be using the opposite orientation. Our strategy to construct a basis of $L ^2 \mathcal{H} ^2 (A _{ K -1 } ) $ will be therefore to look for self-dual 2-forms and impose closure and square-integrability.

The forms $\{\OO ^I \} $ that we are going to construct, see equation (\ref{oi}), have been found before \cite{Ruback:1986ty}. We reproduce them here for two reasons: On one hand to provide an explicit derivation and a proof of the fact that they form a basis of $L ^2 \mathcal{H} ^2 (A _{ K -1 } ) $, both of which are not available in the literature; on the other hand to clarify their topological origin: As we will see $[\OO ^I]$ is the Poincar\'e dual of a 2-cycle $[\SS _I ] $ naturally emerging in the HHM compactification of $A _{ K -1 }$.

The metric of an $A _{K-1} $ ALF gravitational instanton is of  Gibbons-Hawking form,
\begin{equation}
\label{akmetric} 
\mathrm{d} s ^2 
=V ( \mathrm{d} r ^2 + r ^2 \mathrm{d} \Omega ^2 )+ V ^{-1} (\mathrm{d} \psi + \alpha ) ^2,
\end{equation} 
where $\mathrm{d} \Omega ^2 =\mathrm{d} \theta ^2 + \sin ^2 \theta\, \mathrm{d} \phi ^2 $, $(r, \theta,\phi)$ are spherical coordinates in $\mathbb{R}  ^3 $, $\psi\in[0, 2 \pi ) $ is an angle, $\alpha$ is a 1-form locally such that $\mathrm{d} \alpha =* _3 \mathrm{d}V $, $* _3 $ being the Hodge operator with respect to the Euclidean metric on $\mathbb{R}  ^3 $. The function $V$ is given by
\begin{equation}
\label{vak} 
V =1+ \frac{1}{2} \sum _{I=1 }^{ K }  \frac{1}{\norm{p-p^I}},
\end{equation} 
where $\norm{ \cdot } $ is the Euclidean norm in $\mathbb{R}  ^3$ and the points  $\{ p^I \} $ are $K$ distinct points in $\mathbb{R}  ^3 $,  fixed points of the $U (1) $ isometry generated by the Killing vector $\partial / \partial \psi $ known as NUTs.
The metric can be smoothly extended to the points $\{p ^I \} $ provided that they are all distinct.
We denote by $(p ^I _x , p ^I  _y , p ^I _z ) $ the Cartesian coordinates of $p ^I $.
For future convenience let us write
$V =1 + \sum _{  I =1 }^K V ^I$, $\alpha =\sum _I \alpha ^I $, with $\mathrm{d} \alpha ^I =* _3 \mathrm{d} V ^I $.

If $(r ^I , \theta ^I , \phi ^I )$ are spherical coordinates centered at $p ^I $, the form $\alpha ^I $ is given locally, up to addition of a closed 1-form, by the expression
\begin{equation}
\alpha ^I = \frac{1}{2} \cos \theta ^I\,  \mathrm{d} \phi ^I
\end{equation} 
and has the usual Dirac string singularity along the surface $x =p ^I _x $, $y =p ^I _y  $, with $x, y, z $ Cartesian coordinates on $\mathbb{R}  ^3 $. The singularity can be avoided by defining
the two gauge potentials
\begin{equation} 
\label{aans} 
\begin{split} 
\alpha ^I _{ (N) } = \frac{1}{2} \left( \cos \theta ^I -1 \right) \mathrm{d} \phi ^I &\text{\quad for $\theta ^I \neq \pi $, } \\
\alpha ^I _{ (S) } = \frac{1}{2} \left( \cos \theta ^I +1 \right) \mathrm{d} \phi ^I &\text{\quad for $\theta ^I \neq 0$.} 
\end{split} 
\end{equation} 
Note that $\mathrm{d} \alpha ^I _{ (N) } = \mathrm{d} \alpha ^I _{ (S) } $.
Both $\alpha ^I _{ (N) } $ and $\alpha ^I _{ (S) } $ are not defined at $p ^I $ where the angular coordinates are ill-defined.

Introduce the orthonormal coframe
\begin{equation}
\begin{split} 
e ^i &= \sqrt{ V }\, \mathrm{d}x ^i, \quad i =1,2,3,\\
e ^4 & = \frac{1}{\sqrt{ V }} (\mathrm{d} \psi + \alpha ).
\end{split} 
\end{equation} 
We choose the orientation opposite to the one induced by the hyperk\"ahler structure, so our canonical volume element is
\begin{equation}
\label{orak} 
\mathrm{vol} _{ A _{ K-1 } } 
=e ^4 \wedge e ^1 \wedge e ^2 \wedge e ^3
=V ^{-1} r ^2 \sin \theta \, \mathrm{d} \psi \wedge \mathrm{d} r \wedge \mathrm{d} \theta \wedge \mathrm{d} \phi .
\end{equation} 

Start with the self-dual ansatz
\begin{equation}
\label{fft} 
\OO 
=a   _i  \left( e ^4 \wedge e ^i +\frac{1}{2}   \epsilon _{ ijk }e ^j \wedge e ^k \right) 
= a _i \left[ ( \mathrm{d} \psi + \alpha )\wedge \mathrm{d} x ^i +\frac{V}{2}  \epsilon _{ ijk }\,\mathrm{d} x ^j \wedge \mathrm{d} x ^k \right] ,
\end{equation} 
with $a _i $ satisfying $\partial _\psi  a _i   =0 $, $i =1, 2 ,3 $, and impose the closure condition
\begin{equation}
\label{clo} 
\begin{split}
\mathrm{d} \OO  &
=0
= \mathrm{d}  a  _i \wedge (\mathrm{d} \psi + \alpha )\wedge \mathrm{d}  x ^i + a _i\,  \mathrm{d} \alpha \wedge \mathrm{d} x ^i 
+ \frac{1}{2}  \mathrm{d} ( a _i  V )\wedge  \epsilon _{ i jk} \,\mathrm{d} x ^j  \wedge \mathrm{d} x ^k.
\end{split}
\end{equation} 
Since $\partial _\psi  a _i   =0 $, $ \mathrm{d} a _i \wedge \mathrm{d} x ^i =\mathrm{d} \left( a _i \mathrm{d} x ^i \right)  $ must vanish hence, since $H ^1 _{ \mathrm{dR} } ( A _{ K-1 }  ) =0 $,  $a _i  =\partial _i A  $ for some function $A$.
Using  $ \mathrm{d} \alpha =* _3 \mathrm{d} V $  (\ref{clo}) reduces to
\begin{equation}
\mathrm{d}  * _3 \mathrm{d}  ( A   V ) =0,
\end{equation} 
that is  
$ A =f/ V  $
  with $V$ given by (\ref{vak}) and $f$ harmonic with respect to the 3D Euclidean Laplacian.
Therefore
\begin{equation} 
\label{oon} 
\OO
= ( \mathrm{d} \psi + \alpha )\wedge \mathrm{d} A +V  * _3 \mathrm{d} A
\end{equation} 
is closed and self-dual hence  harmonic.

We still need to impose square-integrability. We have
\begin{equation}
\begin{split} 
\OO \wedge * \OO  &
=2 \sum _{ i =1 }^3 ( \partial _i A )^2 \ \mathrm{vol} _{ A _{ K-1 } }. \ok
\end{split} 
\end{equation} 
For large $r$
\begin{equation}
\mathrm{d} A
= \mathrm{d} (f/V)
= \mathrm{d} f  \left( 1 - \frac{k}{2r} + O \left( r ^{ -2 } \right) \right) + f  \cdot O \left( r ^{-2} \right),
\end{equation} 
$\mathrm{vol} _{ A _{ K -1 } }= O ( r ^2) $, therefore  $\OO$ is square-integrable if
  $f$ is either constant or decays at infinity like $1/r$ or faster. For $f=-c$ constant  we get
\begin{equation}
\left. \OO \right | _{ f =-c}
=c \left[ \mathrm{d} V ^{-1} \wedge (\mathrm{d} \psi + \alpha )+ V ^{-1} * _3 \mathrm{d} V  \right].
\end{equation} 

By Liouville's theorem a harmonic function bounded from above or from below and globally defined on $\mathbb{R}  ^n $ must be constant, hence to get a non-constant $f$ we must allow for poles.
In order for  $\OO $ to remain smooth the poles of $f$  must be located at the NUTs positions.
A harmonic function with the required decay at infinity and with poles at the points $p ^I $ must be of the form $\sum _ I  c _I V ^I $, with $c _I $ arbitrary constants and $V ^I =1/(2 r ^I) $. 
Since (\ref{oon}) depends linearly on $f$,
\begin{equation} 
\label{oi} 
\left. \OO \right | _{ f = \sum c _I \, V ^I } = \sum  c _I \, \left. \OO \right | _{f =V ^I }
\end{equation} 
and we only need to consider the 2-forms
\begin{equation}
\begin{split} 
\OO ^I &
\equiv \left. \OO \right | _{ f =V ^I /( 2 \pi ) }\\ &
=\frac{1}{2 \pi }\partial    _i\left( \frac{V ^I }{V } \right)  \left( e ^4 \wedge e ^i +\frac{1}{2}   \epsilon _{ ijk }e ^j \wedge e ^k \right) \\ &
=\frac{1}{2 \pi } \left[ ( \mathrm{d}  \psi + \alpha ) \wedge \mathrm{d} \left( \frac{V ^I }{V } \right) +  V * _3 \mathrm{d} \left( \frac{V ^I }{V } \right) \right] ,\ok
\end{split} 
\end{equation} 
where the normalisation factor $( 2 \pi )^{-1} $ has been chosen for future convenience. Note that $\OO ^I = \mathrm{d} \oo ^I $, with
\begin{equation}
\label{ooi} 
\begin{split} 
\oo ^I &
=\frac{1}{2 \pi } \left( \alpha ^I - \frac{V ^I }{V } \left( \mathrm{d} \psi + \alpha \right) \right)
=\frac{1}{2 \pi } \left[ \alpha ^I \left( 1- \frac{V ^I }{V } \right) -\frac{V ^I }{V } \left( \mathrm{d} \psi+  \sum _{ J \neq I } \alpha ^J  \right) \right]
. \ok
\end{split} 
\end{equation} 
However $\OO ^I $ is not exact as $\oo^I $ is only locally defined. In fact near $p ^I $
\begin{equation} 
\oo ^I 
=-\frac{1}{2 \pi }
 \left( \mathrm{d} \psi + \sum _{ J \neq I } \alpha ^J  
  \right)
+ O \left(r ^I  \right)
\end{equation} 
and $\mathrm{d} \psi $ is not well-defined at $p ^I $.

We can recover the case $f=\mathrm{const} $ by summing  over all NUTs:
\begin{equation}
\label{ooinf} 
\OO ^\infty 
\equiv\sum _{I =1 }^K \OO ^I 
=\frac{1}{2 \pi  }  \left[ \mathrm{d} V ^{-1} \wedge (\mathrm{d} \psi + \alpha )+ V ^{-1} * _3 \mathrm{d} V  \right]
=\left. \OO \right | _{f=-1/(2 \pi )} .  \ok
\end{equation} 
The notation $\infty $ in $\OO ^\infty $ is due to the fact that, as follows from equation (\ref{pdinf}) below, $[\OO ^\infty ] $ is the Poincar\'e dual of $[\SS _\infty ] $.
Note that $ \OO ^{ \infty } =\mathrm{d} \oo ^{ \infty } $, with
\begin{equation} 
\oo ^{ \infty }
=\frac{V ^{-1}}{2 \pi }    (\mathrm{d} \psi + \alpha ).
\end{equation} 
The form $\oo ^\infty $ is globally defined and vanishes at the NUTs positions because of the factor $V ^{-1} $. In fact $2 \pi \oo ^\infty $  is the metric dual of the Killing vector field $\partial  _\psi $.
However $\oo ^\infty $ is not square-integrable as 
\begin{equation} 
\norm{ \oo ^\infty} ^2 
= \int _{  A _{ K -1 } } \oo ^\infty \wedge * \oo ^\infty 
= \int _{  A _{ K -1 } }\!\!\!\! \frac{\mathrm{vol} _{ A _{ K -1 } }}{2 \pi V } 
=4 \pi  \int _0 ^\infty r ^2\, \mathrm{d} r,
\end{equation} 
therefore $ \OO ^{ \infty } $ is exact but not $L ^2 $-exact.

In order to verify that $\{ \OO ^I \} $, $ I =1, \ldots , K  $, is a basis of $L ^2 \mathcal{H} ^2 (A _{ K -1 } ) $ we shall first check that any subset of $K -1 $ elements is a basis of $H ^2 _{ \mathrm{dR}} (A _{ K -1 } ) $ by computing the period matrix.

Let us first describe a convenient basis of $H _2 (A _{ K -1 }; \mathbb{Z}  )$.
An ALF $A _{ K-1 }$  gravitational instanton retracts onto a configuration of 2-spheres intersecting according to minus the Cartan matrix of the $A _{ K -1 } $ Lie algebra.\footnote{
The usual choice of orientation, opposite to (\ref{orak}), would give the plus sign. The Cartan matrix of the Lie algebra $A _{ K -1} $ is tri-diagonal with 2 on the main diagonal and $-1 $ above and below it.
}
It is possible to represent the homology classes of these 2-spheres by minimal area embedded 2-surfaces having the topology of a 2-sphere \cite{Sen:1997tw}.
They are constructed as follow. Take the line segment connecting two distinct NUTs and passing through no other NUT. Above each point of this line there is a circle, which collapses to zero radius at both ends of the segment. Hence the resulting surface is topologically a 2-sphere. By considering the  induced metric one can check that, apart from a constant factor, the area of this surface is equal to the length of the segment connecting the two NUTs. Since straight lines minimise Euclidean length, the surface has minimal area, at least among surfaces having the same topology. Some properties of these surfaces have been studied in \cite{Franchetti:2013ib}.

\begin{figure}[htbp]
\begin{center}
\includegraphics[width=0.6\textwidth]{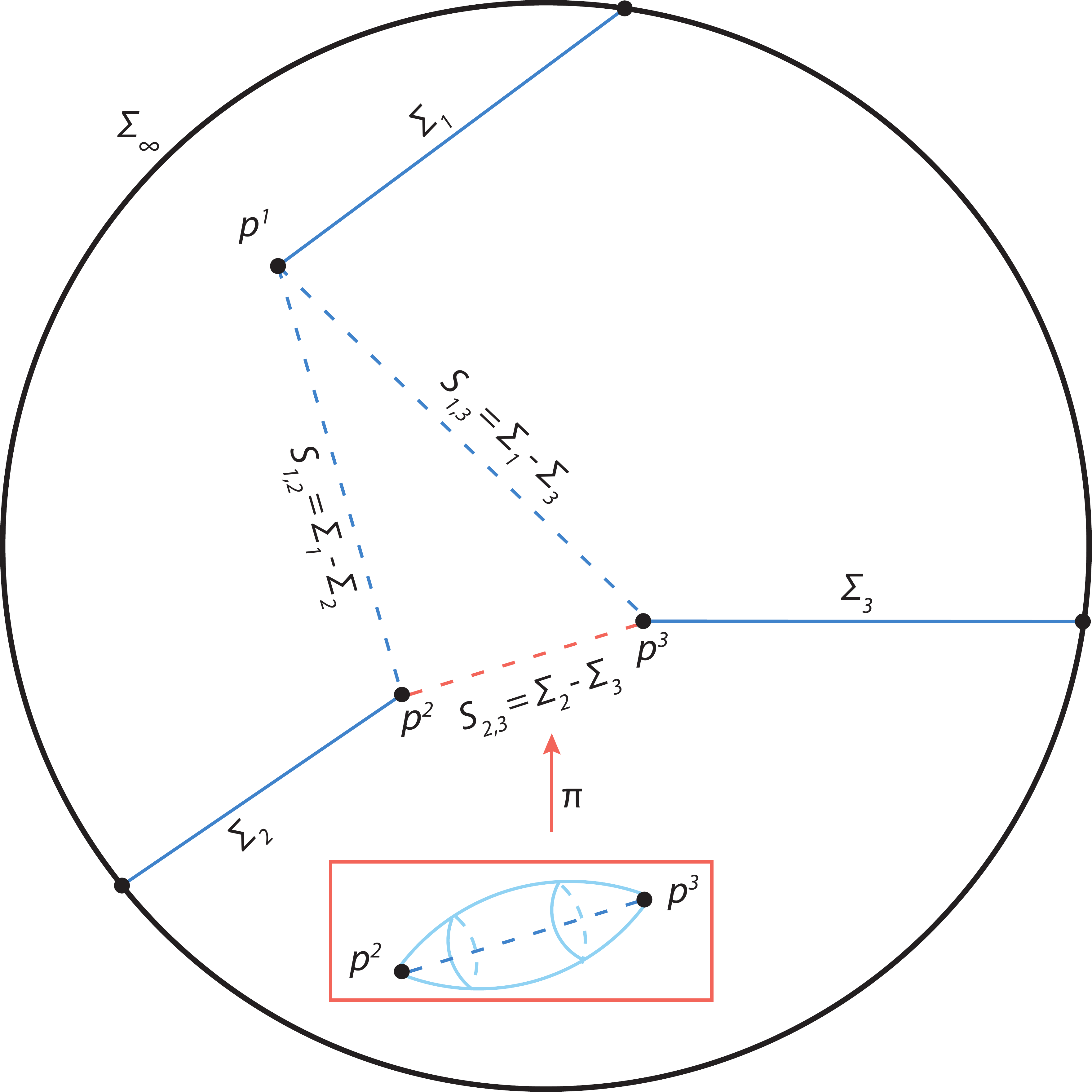}
\caption{Homology of $A _2 $ and $X _{ A _2 } $. Black filled dots correspond to NUTs. The surface at infinity $\SS_{ \infty } $  can be thought as a 2-sphere worth of NUTs. Connecting any two NUTs by a line segment gives a 2-cycle. The angular coordinate $\psi$ has been suppressed so  2-cycles appear as line segments, the fibration by circles is shown in the box.  Continuous lines correspond to the 2-cycles $\SS _1 $, $\SS _2 $, $\SS _3 $ generating $H _2 (X _{ A _2 }; \mathbb{Z}  )$, dashed lines correspond to 2-cycles $\S_{1,2} $, $S_{1,3} $, $\S_{2,3} $ generating $H _2 (A _2; \mathbb{Z}  )$.}
\label{xa2hom}
\end{center}
\end{figure}

If $p ^I $, $p ^J $ are two NUTs such that the line segment connecting them passes through no other NUT,  denote by $\S_{  I, J }$ the associated minimal area surface, and by $[\S _{ I , J }] $ the corresponding  homology class  in $ H_2  (A _{ K-1 }; \mathbb{Z}  ) $. The construction is illustrated in figure \ref{xa2hom} for the case $K =3 $. 
Consider the integral
\begin{equation}
\label{dhfg} 
\int _{ \S_{ I, J }} \OO ^K .
\end{equation} 
If we use spherical coordinates $(r, \theta , \phi )$ such that the line through $p ^I $, $p ^J $ has constant angular coordinates $(\theta = \theta _0 ,\phi = \phi _0 )$, we can parametrise $\S_{ I , J } $ as
\begin{equation}
\S_{ I , J }
=\{ (r, \psi, \theta _0 , \phi _0  )\in A _{ K -1 }: p ^I _r \leq r \leq p ^J _r , \psi \in[0, 2 \pi )   \},
\end{equation} 
where $p ^I _r $ ($p ^J _r $) is the $r$-coordinate of $p ^I $ ($p ^J $).
On  $\S_{ I , J }$ we take the orientation  $\mathrm{d} r \wedge \mathrm{d} \psi  $, with  $r $ increasing in the direction of $p ^J $.

Recall that $\OO ^K =\mathrm{d} \oo ^K $, with $\oo ^K $ given by (\ref{ooi}).
If $ K \neq I $, $ K \neq J $ then $p ^K \notin \S_{ I , J }$ and the term $(V  ^K/ V )( \mathrm{d} \psi + \alpha ) $ is well defined on $\S_{ I , J }$.
If $\alpha ^K $ is globally defined on $\S_{I,J} $ then $\OO ^K $ is exact and the integral vanishes, otherwise we can break the integration region into two parts in each of which $\OO ^K $ is exact. Define
\begin{equation}
\oo ^K _{ (N) }= \frac{1}{2 \pi } \left( \alpha ^K _{ (N) }- \frac{V ^K }{V } (\mathrm{d} \psi + \alpha ) \right) ,\qquad 
\oo ^K _{ (S) }= \frac{1}{2 \pi } \left( \alpha ^K _{ (S) }- \frac{V ^K }{V } (\mathrm{d} \psi + \alpha ) \right) ,
\end{equation}  
with $ \alpha ^K _{ (N) }$, $\alpha ^K _{ (S) }$ given by (\ref{aans}) and let
\begin{equation}
\S_{ I , J (N)}
=\{(r, \psi )\in \S_{ I , J } : \OO ^K = \mathrm{d} \oo ^K _{ (N) } \}, \qquad 
\S_{ I , J(S)}
=\{(r, \psi )\in \S_{ I , J } : \OO ^K = \mathrm{d} \oo ^K _{ (S) } \}.
\end{equation} 
The surfaces $\S_{ I , J (N)} $, $\S_{ I , J(S) } $ have a circle parametrised by $\psi$  as the common boundary, but with opposite induced orientation, hence (\ref{dhfg}) reduces to
\begin{equation}
\int _{S ^1 } (\alpha ^K _{ (N) }-\alpha ^K _{ (S) } )=0
\end{equation} 
as $\alpha ^K _{ (N) }= 0=\alpha ^K _{ (S) }$ when restricted to this circle.

If $ K =I $, $\epsilon >0 $,  let $\S_{ I , J } ^\epsilon=\{(r, \psi )\in S ^{ I , J }: p ^I _r  + \epsilon \leq r \leq p ^J _r \} $. Then $ \partial \S_{ I , J } ^\epsilon$ is a small circle of radius $\epsilon$ and the induced boundary orientation is $- \mathrm{d} \psi $.
On $\S _{I,J} ^\epsilon $ the form $ \OO ^I $ is exact, hence
\begin{equation}
\label{compp} 
\begin{split} 
\int _{ \S _{  I, J }  } \OO ^I &=
\lim _{ \epsilon \rightarrow 0 } \int _{ \S _{  I, J }^\epsilon } \OO ^I 
=\frac{1}{2 \pi }\lim _{ \epsilon \rightarrow 0 } \int _{ \partial S _{  I, J } ^\epsilon } \left( \alpha ^I - \frac{V ^I  }{V } (\mathrm{d} \psi + \alpha ) \right) 
=\frac{1}{2 \pi }\lim _{ \epsilon \rightarrow 0 }\int _{ S ^1  _\epsilon } \frac{V ^I  }{V } \mathrm{d} \psi \\ &
= \frac{1}{2 \pi } \cdot  2 \pi  \lim _{p \rightarrow p ^I  } \frac{V ^I }{V }
=1.
\end{split} 
\end{equation} 

The case $K = J $ is obtained from  $K = I $ by an orientation reversal, hence we obtain the period matrix
\begin{equation}
\int _{ \S_{ I, J }} \OO ^K = \delta _I ^K - \delta _J ^K
\end{equation} 
which has maximal rank $ K -1 $. Therefore any $K -1 $ elements of $\{ [\OO ^I] \} $ form a basis of $H ^2 _{ \mathrm{dR} }(A _{ K -1 } )$. As we noticed before $\OO ^\infty =\sum _I \oo ^I $ is exact but not $L ^2 $-exact hence $\{ \OO ^I \} $, $I =1 , \ldots , K $, is a basis of $L ^2 \mathcal{H} ^2 (A _{ K -1 }) $.

Label the NUTs so that the line segment from $p ^I $ to $p ^{ I+1} $ passes through no other NUT.
We can  construct a basis $\{[ \O^{I,I+1}]\} $ of $H ^2 _{ \mathrm{dR}  } (A _{ K -1 } ) $ consisting of square-integrable harmonic forms such that $[\O ^{  I,I+1}]$ is the Poincar\'e dual of $[\S_{  I , I+1 }] $: Define
$\O ^{I,I+1} = \OO ^{ I + 1 }- \OO ^I$, $ I =1 , \ldots , K -1 $,
then
\begin{equation}
\label{oo2} 
\int _{ \S_{J, J + 1 }  } \O ^{I,I+1}
= \delta ^{I-1} _{J}+ \delta ^{ I + 1 }_{ J } -2 \delta ^I _J .\ok
\end{equation} 

Since the HHM compactification of an ALF gravitational instanton  is obtained by collapsing the fibre above each point of $\SS _r $ as $r \rightarrow \infty $, $\SS_{ \infty }$ can be thought as a 2-surface, in the case of $A _{ K -1 }$ a 2-sphere, worth of NUTs, see figure \ref{xa2hom}.
Therefore any half-line connecting a NUT $p ^I $ to a point of the 2-sphere at infinity $\SS_{ \infty }$ defines a surface $\SS_I $, topologically a 2-sphere,  representing a homology class $[\SS_I ]\in H _2 ( X _{ A _{ K -1 } }; \mathbb{Z}  )$. We take on $\SS_{ I }$ the orientation $\mathrm{d} r \wedge \mathrm{d} \psi $, with $r$ increasing toward $\infty $.
A computation similar to  (\ref{compp}) gives
\begin{equation}
\label{pdinf} 
\int _{ \SS_I } \OO ^J = \delta ^{ J } _I,\ok
\end{equation} 
therefore $[\OO ^I] $ is the Poincar\'e dual of $[\SS_I] $ in $X _{ A _{ K-1 } }$ and $\{ [\OO ^I]  \} $, $I =1 , \ldots, K$, is a basis of $H ^2 _{ \mathrm{dR} }(X _{ A _{ K -1 } } )$. Note that (\ref{pdinf}) also implies that surfaces $\SS_I $, $\tilde \SS_I $ obtained by connecting $p ^I $ to two different points of $\SS_{ \infty } $ are homologous.
It is interesting to notice that $[\OO ^\infty ] $, with $\OO ^{ \infty }$ given by (\ref{ooinf}), 
is Poincar\'e dual in $X _{ A _{ K -1 } } $ to the 2-cycle $\sum _{I =1 }^K [\SS_I] =[\SS_\infty]$.

Denote by $ S \cdot _{ X} S ^\prime  $ the intersection number of $S$ and $S ^\prime $ inside the space $X$. By definition of Poincar\'e dual we have
\begin{align} 
\S_I  \cdot_{ A _{ K -1 } } \S_J   &
=\int _{ A _{ K-1 }  } \O ^{I,I+1} \wedge \O^{J,J+1}  
= \delta ^{I-1} _{  J }+ \delta ^{ I + 1 }_{ J } -2 \delta ^I _J , \ok\\
\SS_I \cdot _{ X_{A _{ K -1} } }\SS_J &
=\int _{ X _{ A _{ K-1 }  } } \OO ^I \wedge \OO ^J  
 =\delta _I ^J ,\ok
\end{align} 
which can be also checked by direct computation.

To summarise:  $\{\OO ^I \} $ is a basis of   $ L ^2 \mathcal{H} ^2 (A _{ K -1 }) $, $\{ [\SS_I] \} $ is a basis of $H _2 (X _{ A _{ K -1 }} ; \mathbb{Z}  )$
and $[\OO ^I ] $ is the  Poincar\'e dual of $[\SS ^I ] $;
$\{ [\O ^{ I , I +1}] \} $ is a basis of $H ^2 _{ \mathrm{dR} } (A _{ K -1 }) $ 
with $\O^{I,I+1} $ a harmonic square-integrable 2-form, $\{[\S_{I, I + 1 }]\} $ is a basis of $H _2 (A _{ K -1 }; \mathbb{Z}  ) $ with $\S_{I,I+1} $ a minimal area embedded surface, and $[\O^{I,I+1} ]$ is the Poincar\'e dual of $[\S_{I, I + 1 }]$. The homology and cohomology of $A _{ K -1 }$, $X _{ A _{ K -1 } }$ are related by the equations $ [\S_{I, I +1 }]=[\SS_I] - [\SS_{ I + 1 }] $, $ [\O ^{ I , I + 1 }]=[ \OO ^{I+1}]- [\OO ^I ] $. Note that  greek letters identify objects naturally related to $X _{A _{ K -1 }} $ and latin letters identify object naturally related to $A _{ K -1 }$. The (cohomology class of the) 2-form $\OO ^{ \infty }\in L ^2 \mathcal{H} ^2 (A _{ K -1 } ) $ which is singled out by the fact of being exact (but not $L ^2 $-exact) is Poincar\'e dual in $X _{ A _{ K -1 } } $ to the cycle at infinity $[\SS_{ \infty }]$. See also figure \ref{xa2hom} for a pictorial representation of these results.

We now come to the $D _K $ family.
The construction of the exact metric on $D _K $ is quite involved \cite{Cherkis:2005ba},
however there is an asymptotic approximation of Gibbons-Hawking form \cite{Chalmers:1999vb},
\begin{equation}
\label{dkmetric} 
\mathrm{d} s ^2 
= V (\mathrm{d} x ^2 + \mathrm{d} y ^2 + \mathrm{d} z ^2 )+ V ^{-1} (\mathrm{d} \psi + \alpha ) ^2 ,
\end{equation}  
with $\alpha$ locally such that $\mathrm{d}  \alpha =* _3 \mathrm{d} V  $, and
\begin{equation}
\label{vvdk} 
V =1 - \frac{2}{\norm{p}} + \frac{1}{2} \sum _{ I =1 }^K \left( \frac{1}{\norm{p-p ^I} } + \frac{1}{\norm{p + p ^I} } \right),
\end{equation} 
where $ \{p ^I  \} $ are $K$ distinct points in $\mathbb{R}  ^3 $. There is the $\mathbb{Z}    _2 $ identification
\begin{equation}
\label{i1} 
\psi \sim - \psi , \quad
\theta \sim \pi - \theta , \quad
\phi \sim \phi + \pi.
\end{equation}
For future convenience write 
\begin{equation}
 V =1- \frac{2}{\norm{p}} + \sum _{ I =1 }^k  V ^{ I + }   + \sum _{ I =1 }^k  V ^{ I - } ,
\end{equation}
with $V ^{ I \pm }= (2\norm{p \mp p ^I }) ^{-1} $ and define $\alpha ^{ I \pm } $ via  $ \mathrm{d} \alpha ^{ I \pm }  =* _3 \mathrm{d} V ^{ I \pm } $.

As long as we work with the approximate metric (\ref{dkmetric}) we can proceed as we did for $A _{ K -1 } $.
By looking for self-dual, square-integrable harmonic forms on $ A _{ K -1 } $ we obtained the expression
\begin{equation}
\OO
= ( \mathrm{d} \psi + \alpha )\wedge \mathrm{d} (f/V) +V  * _3 \mathrm{d} (f/V) 
\end{equation} 
with $f$ either constant or given by a superposition of poles located at the NUTs. Since we did not make use of the detailed form of $V$, the expression is  valid also for $D _K $, but we now need to take into account the $\mathbb{Z}  _2 $-identification (\ref{i1}) under which $V$ is even,  $ \psi$ and $\alpha$ are odd. In order for $\OO $ to be invariant   $f$ needs to be odd, hence we cannot have  $f= \mathrm{const}  $  or containing a pole at the origin and other poles of $f$ must appear in the odd combination $ V ^{ I +} - V^{I-}$. Therefore we obtain $K$ harmonic square-integrable 2-forms $ \{ \OO ^I \} $ by taking $f =  ( V ^{ I +} - V^{I-})/(4 \pi ) $ with the normalisation factor chosen for future convenience.

We will now show that $\{[\OO ^I] \} $ is a basis of $H ^2 _{ \mathrm{dR}} (D _K ) $ by looking at the period matrix.
We first need to construct a convenient basis of $H _2 (D _K ; \mathbb{Z}  )$.
To a line segment connecting $p ^I $ to $p ^J $ and passing through no other NUT  is associated a minimal area 2-cycle $\S_{ I,J } $ constructed as before.
Label the points $\{ p ^I \} $ in (\ref{vvdk}) so that the line segment between $p ^I $ and $p ^{ I +1 } $ passes through no other NUT, and so does  the line segment between $p ^{ K -1 } $ and $- p ^K $. Denote by $\S_{ I, I+1 } $ the 2-cycle obtained by connecting the NUT $p ^I $ to the NUT $p ^{ I+1} $ and by $\S_{ K-1, -K} $ the 2-cycle obtained by connecting the NUT at $p ^{ K-1 } $ to the one at $-p ^K $. Then $\{ [\S_{ 1,2}], \ldots , [\S_{ K -1 , K }], [\S_{ K -1, -K}]\} $
is a basis of $H _2 (D _K; \mathbb{Z}   )  $ with intersection matrix minus the Cartan matrix of the Lie algebra $D _K $ whose representatives are (approximately) minimal area embedded surfaces.
Calculations similar to those we did for $A _{K-1} $ show that 
\begin{equation}
\int _{ \S_{ I, J }} \OO ^K = \delta _I ^K - \delta _J ^K ,
\end{equation} 
hence
$\{ \OO ^I \} $ is an approximate basis of $L ^2 \mathcal{H} ^2  (D _K ) $, and  $\{ [\OO ^I] \} $ an approximate basis of  $H ^2 _{ \mathrm{dR}} (D _K ) $.

We can also construct harmonic square-integrable representatives of the Poincar\'e duals of the 2-cycles $[\S_{I,J}] $: The  combination $[\O ^{ I, I+1}]=[\OO ^{  I + 1 }]-[ \OO ^I ]$, $I=1, \ldots , K-1 $,
is the Poincar\'e dual of $[\S_{I,I+1}]$, and
$[\O ^{K-1, -K }]=- ([\OO ^K] +[ \OO ^{ K + 1 }] ) $ is the Poincar\'e dual of $[\S_{K-1, -K}$].
Finally, if $\SS_J $ is the surface connecting $p ^J  $ to a point on the surface at infinity then $\{[\SS _I ]\} $, $I =1 , \ldots, K $, is a basis of $H _2 (X_{D _K} ; \mathbb{Z}  ) $ with $[\SS _I ] $ Poincar\'e dual to $[\OO ^I ] $ since
\begin{equation} 
\int _{ \SS_{ J  } } \OO ^{ I } = \delta ^I _J.
\end{equation} 

As we can see, the main differences between  the Hodge cohomology of $A _{ K -1 } $ and $D _K $ stem from the fact that the surface at infinity in $D _K $ does not contribute to the middle dimension homology of $X _{ D _K }$. In turn, this can be tracked down to the  
$\mathbb{Z}  _2 $-identification (\ref{i1}) featured by $D _K $ but not by $A _{ K -1 } $ as a consequence of the different properties of the underlying topological manifolds.

\appendix
\section{The topology of large $r$ hypersurfaces}
\label{ga} 
Outside a compact set the topology of an ALF gravitational instanton is that of $( \mathbb{C}  ^2 \setminus \{0\}) / \Gamma $ with $\Gamma $ a finite subgroup of $SU (2) $. Let $(z ^1 , z ^2 )$ be complex coordinates on $\mathbb{C}  ^2 $. 

For ALF $A _{ K -1 } $,  $\Gamma =\mathbb{Z}  _K $. The action of $\mathbb{Z}  _K $ on $\mathbb{C}  ^2 $ is generated by
\begin{equation}
\label{aka} 
(z ^1 , z ^2 ) \mapsto  \exp \left( i\,2 \pi /K \right) (z ^1 , z ^2 ).
\end{equation} 
In terms of spherical coordinates $r\in(0, \infty )$, $\theta \in[0, \pi ] $, $ \phi \in[0, 2 \pi )$, $\x \in[0,4 \pi )$  (\ref{aka}) becomes
\begin{alignat}{4}
&r \mapsto r, \quad &\theta \mapsto \theta, \quad & \phi \mapsto \phi, \quad & \x \mapsto \x + 4 \pi /K.
\end{alignat} 
Therefore $\mathbb{Z}  _K $ only acts on the fibres and a hypersurface of large $r$ is $S ^3 $  for $K =1 $ (Hopf fibration) and the lens space $L(K,1) $, a $U (1) $ bundle over $ S ^2 $ with Chern number $K$, for $K >1 $.\footnote{
The angle $\psi\in[0, 2 \pi ) $ appearing in the asymptotic $A _{ K  -1 }$ metric (\ref{akmetric}) is related to $\x $ by  $\psi = K  \alpha /2 $.}

For $D _K $, $K \geq 3 $,  $\Gamma =D ^\ast _{ K -2} $, the binary dihedral group of order $4(K-2) $.
The group $D ^\ast _K $ has presentation
\begin{equation}
D^*_{K}
=\langle a,b : a^{2K}=e, b^2=a^{K}, b\,a\,b^{-1}=a^{-1} \rangle.
\end{equation} 
The action of $D ^\ast _{ K -2 } $ on  $\mathbb{C}  ^2 $ is generated by
\begin{alignat}{2}
\label{dka} 
&a :&\quad  &(z ^1 , z ^2  )\mapsto \exp \left( i\, \pi/(K-2)  \right) (z ^1 , z ^2 ),\\
\label{dkb} 
&b :&\quad &(z ^1 , z ^2  )\mapsto i (\bar z ^2 ,- \bar z ^1 ).
\end{alignat} 
or, in terms of the spherical coordinates $(r, \theta, \phi , \x) $,\footnote{
The angle $\psi\in[0, 2 \pi ) $ appearing in the asymptotic $D _K $ metric (\ref{dkmetric}) is related to $\xx $ by $\psi =(K-2) \xx$.
}
\begin{alignat} {5}
\label{dkid} 
&a :&\quad & r \mapsto r , &\quad & \theta \mapsto \theta, &\quad & \phi \mapsto \phi, &\quad &\x \mapsto  \x + 2 \pi /( K -2 ),\\
&b :&\quad & r \mapsto r , &\quad & \theta \mapsto \pi -\theta, &\quad & \phi \mapsto \pi + \phi, &\quad &\x \mapsto -\x.
\end{alignat} 
For $K \geq 3 $ a large $r$ hypersurface has therefore the topology of $S ^3 / D ^\ast _{ K -2 } $.
Note that while $D ^\ast _1 = \mathbb{Z}  _4 $, its action on ALF $ D _2 $ is different from the $\mathbb{Z}  _4 $-action on ALF $ A _3 $.

ALF  $D _0 $, the moduli space of charge 2 centred $SU (2) $ monopoles, is modded out by the  transformations (see \cite{Gibbons:1986wu} where the angle we denote by $\xx $ is denoted by $\psi$)
\begin{alignat}{5}
&I  _1 :&\quad  &r \mapsto r, &\quad &\theta \mapsto \pi - \theta,&\quad  &\phi \mapsto \pi + \phi, &\quad & \xx \mapsto - \xx, \\
&I _2 :& \quad  &r \mapsto r, &\quad &\theta \mapsto \pi - \theta , &\quad &\phi \mapsto \pi + \phi , &\quad & \xx \mapsto \pi - \xx,\\
\label{i3} 
&I _3 :& \quad  &r \mapsto r, &\quad &\theta \mapsto \theta, &\quad &\phi \mapsto \phi , &\quad & \xx \mapsto \pi + \xx,
\end{alignat} 
with $(r, \theta , \phi )$ as before and $\xx\in[0, 2 \pi )$.
Note that $I _3 \circ I _1 =I _2 $. Since $I _1 =b$, $I _3 =\left. a \right | _{ K =4} $,  a large $r$ hypersurface in  $D _0 $ has  the same topology,
 $ S^3 / D ^\ast _2 $,  as one in $D _2 $,  but  opposite  orientation \cite{Biquard:2010ig}.

ALF $D _1 $ is modded out by $I _1 $ only. Since $\xx \in[0, 2 \pi )$ we obtain the same identifications as in (\ref{dkid}) for $K =3 $.
A large $r$ hypersurface  has therefore the same topology, $ S^3 / D ^\ast _1 $, as one in $D _3$, but opposite orientation.

ALF $D _2 $ is  the minimal resolution of $ (\mathbb{R}  ^3 \times S ^1)/ \mathbb{Z}  _2 $ \cite{Hitchin:1984wt,Biquard:2010ig}. If $(r, \theta , \phi )$ are spherical coordinates on $\mathbb{R}  ^3 $ and $\xx\in[0, 2 \pi )$ parametrises $S ^1 $, the  $\mathbb{Z}  _2 $-action is generated by
\begin{alignat}{4}
\label{d2a} 
& r  \mapsto r , &\quad &\theta \mapsto \pi - \theta , &\quad \phi \mapsto \pi + \phi, &\quad  &\xx \mapsto 2  \pi -\xx.
\end{alignat} 
The action (\ref{d2a})  is  antipodal  on $\mathbb{R}  ^3 $ but not on $S ^1 $. 
A large $r$ hypersurface has the topology of $(S ^2 \times S ^1) / \mathbb{Z}  _2 $.

Note that the $\mathbb{Z}  _K $-action (\ref{aka}) and the $D ^\ast _{K-2} $-action (\ref{dka}) have the origin of $\mathbb{C}  ^2 $ as their only fixed point,   while the $\mathbb{Z}  _2 $-action (\ref{d2a}) has two fixed points: $( \mathbf{0} , 0) $, $(\mathbf{0} , \pi ) \in \mathbb{R}  ^3 \times S ^1 $, where $\mathbf{0} $ denotes the origin of $\mathbb{R}  ^3 $.

\acknowledgments
G.F.~would like to thank Gabor Etesi, Jos\'e Figueroa-O'Farrill and Bernd Schroers for useful and interesting discussions.

\bibliographystyle{jhep}
\bibliography{harmonic_forms}
\end{document}